\documentclass[conference]{IEEEtran}
\IEEEoverridecommandlockouts
\usepackage{cite}
\usepackage{amsmath,amssymb,amsfonts}
\usepackage{algorithmic}
\usepackage{graphicx}
\usepackage{textcomp}
\usepackage{xcolor}
\def\BibTeX{{\rm B\kern-.05em{\sc i\kern-.025em b}\kern-.08em
    T\kern-.1667em\lower.7ex\hbox{E}\kern-.125emX}}
\begin{document}

\title{FFT and Linear Convolution Implementation with Bit Slicing Multiplier: A Novel Approach\\

}

\author{
\IEEEauthorblockN{Aravind Kumar N}
\IEEEauthorblockA{\textit{Vellore Institute of Technology} \\
\textit{Chennai, Tamilnadu} \\
\textit{aravindkumar.n2020@vitstudent.ac.in}}
\and
\IEEEauthorblockN{Hari Krishna S}
\IEEEauthorblockA{\textit{Vellore Institute of Technology} \\
\textit{Chennai, Tamilnadu} \\
\textit{hari.krishna2020@vitstudent.ac.in}}
\and
\IEEEauthorblockN{Dr. Anita Angeline A}
\IEEEauthorblockA{\textit{Associate Professor} \\
\textit{Vellore Institute of Technology} \\
\IEEEauthorblockA{Chennai, Tamilnadu}
\textit{anitaangeline.a@vit.ac.in}}
}

\maketitle

\begin{abstract}

This paper presents a comprehensive exploration of Fast Fourier Transform (FFT) and linear convolution implementations, integrating both conventional methods and novel approaches leveraging the Bit Slicing Multiplier (BSM) technique. The Bit Slicing Multiplier utilizes Look-Up Tables (LUTs) to execute bitwise operations in parallel, offering efficient arithmetic operations ideally suited for digital signal processing tasks. We extensively investigate the integration of BSM into FFT and linear convolution algorithms, emphasizing its advantages in terms of speed and resource utilization. Additionally, we introduce our own innovative ideas for FFT and convolution algorithms, contributing to the broader discourse on efficient signal processing techniques. Experimental validation of our implementations is conducted using Vivado, a leading FPGA synthesis and implementation tool. Comparative analysis demonstrates the superior performance of our BSM-enhanced approaches, showcasing their potential for real-time signal processing applications. This study not only advances the understanding of FFT and convolution implementations but also highlights the effectiveness of novel techniques like BSM in enhancing computational efficiency in FPGA-based systems.

\end{abstract}

\section{Introduction}
Digital signal processing (DSP) techniques play a pivotal role in a myriad of applications spanning from telecommunications to multimedia processing. Among the fundamental operations in DSP are Fast Fourier Transform (FFT) and linear convolution, which serve as building blocks for a wide range of signal processing algorithms. The efficiency and accuracy of FFT and convolution implementations are paramount, especially in real-time applications where computational resources are constrained.\newline
\newline
Traditional implementations of FFT and convolution algorithms often rely on complex arithmetic operations, leading to significant computational overhead and resource utilization. In recent times, there has been a growing curiosity in exploring alternative techniques to enhance the efficiency of these operations on FPGA-based systems. One such promising approach is the utilization of the Bit Slicing Multiplier (BSM), which offers a novel method for performing arithmetic operations efficiently using Look-Up Tables (LUTs).\newline
\newline
In this paper, we present a thorough investigation into FFT and linear convolution implementations, incorporating our novel approaches leveraging BSM technique. We delve into the underlying principles of BSM and its advantages in terms of speed and resource utilization, highlighting its potential to revolutionize FFT and convolution operations in FPGA-based systems.\newline
\newline
Moreover, we conduct thorough testing and validation using Vivado, a popular FPGA synthesis and implementation tool, to ensure the application meets real-time constraints with effectiveness and reliability. .\newline
\newline
Through comparative analysis and experimental validation, we demonstrate the superiority of BSM-enhanced approaches over traditional methods, showcasing their potential for real-time signal processing applications. This study not only advances the state-of-the-art in FFT and convolution implementations but also underscores the importance of exploring novel techniques like BSM to address the ever-growing demands for efficient signal processing in FPGA-based systems.

\vspace{\baselineskip} 
\section{Literature Review}
\textbf{\textit{}}

\begin{enumerate}

\item{\textbf{Re-configurable Superconducting FFT Processor Using Bit-Slice Block Share Processing Unit[1] :}}

This was helpful in analyzing the Bit - Slicing Technique on a deeper level. The main focus of this paper lies in their novel unit called Bit-Slice Block Share Unit (BSPU), which reduces computational complexity of the radix 2 algorithm to \(\frac{N}{2} {Log_2} {N}\).
BSPU employs a combination of split radix FFT, mixed radix DIT FFT, and bit-serial radix-2 FFT algorithms. It operates on input data sliced into 2-bit segments from a 4-bit source. BSPU finds application in real-time scenarios demanding computations on a larger scale of \( {N^2}\) for efficient processing. The proposed design achieves significant reductions in computation complexity and hardware costs. The performance of the processor is evaluated through latency calculations at 10 GHz, demonstrating improved speed and efficiency compared to conventional semiconductor-based FFT implementations. 
\newline
\item{\textbf{Near-Precise Parameter Approximation for Multiple Multiplications on a Single DSP Block :}}
\\
This research paper was helpful in understanding how to optimize Multiply-Accumulate (MAC) operations on Digital Signal Processor (DSP) blocks. It introduces innovative techniques for manipulating small bit-length framework within a one DSP block, enabling many parameter multiplications with reduced hardware overhead. Their approach not only maintains accuracy but also facilitates significant parameter compression without additional hardware costs. Through exploration of approximation techniques tailored for DSP architectures, the study sheds light on achieving near-precise results while maximizing hardware utilization. The implications extend to various domains reliant on efficient signal processing, offering a promising avenue for enhancing scalability, efficiency, and cost-effectiveness in DSP-based systems.
\\
\item{\textbf{A Flexible-Channel MDF Architecture for Pipelined Radix-2 FFT :}}

Study was helpful in understanding the innovative approach of flexible-channel FC-MDF FFT architecture proposed by Xiao Zhou et al. The paper sheds light on the dynamic adaptation capability of this architecture to accommodate different FFT sizes, leading to optimized resource utilization and improved throughput compared to traditional approaches.
\\
\item{\textbf{Efficient Hardware Implementation of Scalable FFT using Configurable Radix-4/2 :}}
\\
This research paper illuminates the efficacy of leveraging Radix-4/Radix 2*2 FFT algorithms to achieve superior area and performance efficiency, surpassing commercial FFT IP cores. Additionally, the innovative Address Generator architecture showcased in the paper contributes to efficient address mapping and hardware overhead reduction, making it a valuable resource for comprehending state-of-the-art techniques in FFT implementation for wireless and signal processing applications.
\\
\item{\textbf{Implementation of FPGA - Based FFT Convolution :}}
\\
Study was helpful in understanding the development of an efficient overlap-add filter design for linear convolution tasks. The paper delves into the utilization of FFT convolution methods and modified radix-4 architectures to achieve comparable accuracy and throughput with reduced hardware resource requirements compared to designs employing commercial FFT IP cores. Notably, the proposed architecture offers a cost-effective solution tailored for implementing linear convolution on Virtex-5 FPGAs, shedding light on innovative approaches for FPGA-based FFT convolution implementations.
\\
\item{\textbf{An FFT Accelerator Using Deeply-coupled RISC-V Instruction Set Extension for Arbitrary Number of Points :}}
\\
The paper introduces a novel RISC-V ISA extension ploys tailored specifically for efficient FFT processes, incorporating Twelve new instructions. This extension scheme is remarkable for its ability to achieve a substantial reduction in power consumption, requiring less than 16\% compared to utilizing RISC ISA directly for FFT processes. Moreover, the paper outlines the integration of the newly introduced FFT U component into the EX stage of pipeline as an seperate entity. This Unit leverages Single Instruction, Multiple Data technology to execute FFT operation instructions efficiently. These advancements underscore the significance of the proposed approach in enhancing FFT operations while minimizing power consumption, offering valuable insights into optimizing FFT processing within RISC-V architectures.
\\
\item{\textbf{FACCU: Enable Fast Accumulation for High-Speed DSP Systems :}}
\\
The paper presents fast accumulation techniques aimed at boosting processing speed in DSP systems. These FACCU architectures leverage unique features of accumulators to reduce critical path delays and achieve superior performance compared to prior art. The FACCU designs offer scalability and minimal overhead, making them suitable for various high-speed DSP applications
\\
\item{\textbf{Fused Fixed-Point Arithmetic Unit for Radix-4 DIT FFT Implementation :}}
\\
The authors present a novel design approach for computing Radix-4 Decimation in Time Fast Fourier Transform (DIT-FFT) using Fused Arithmetic operations. This method employs Fused Add Subtract (FAS) and Fused Dot Product (FDP) operations to efficiently handle complex fixed-point inputs. Unlike conventional FFT butterfly computations where addition,multiplication and subtraction happen in separate blocks and serially, the approach integrates these operations into fused blocks, resulting in improved throughput, reduced area, and enhanced computational speed. Simulation using Vivado demonstrates that the proposed design occupies Forty Six percent lower area and consumes 2.70 percent lower power compared to conventional radix-4 butterfly computation methods. 
\\
\item{\textbf{Investigation and Vlsi Implementation of Linear Convolution Architecture for FPGA based Signal Processing Applications :}}
\\
The authors propose an architecture leveraging Vedic multiplication techniques to enhance computational speed. This architecture, implemented on Spartan 6 FPGA, demonstrates accelerated convolution operations with real-time verification.
\\
\item{\textbf{Low-Complexity Precision-Scalable Multiply-Accumulate Unit Architectures for Deep Neural Network Accelerators :}}
\\
Recent advancements in Multiply-Accumulate (MAC) unit architectures have garnered significant attention due to their crucial role in accelerating computations in various digital signal processing (DSP) applications.The authors propose low-complexity Precision-Scalable MAC (PSMAC) unit architectures tailored for Deep NN accelerators. By simplifying the architecture and optimizing redundant logic, these PSMAC units achieve substantial savings in area price and power disspation compared to state of-art designs.

\end{enumerate}

\section{Bit Slicing Multiplier(BSM)}

\begin{figure}[h]
    \centering
    \includegraphics[width=1.1\linewidth]{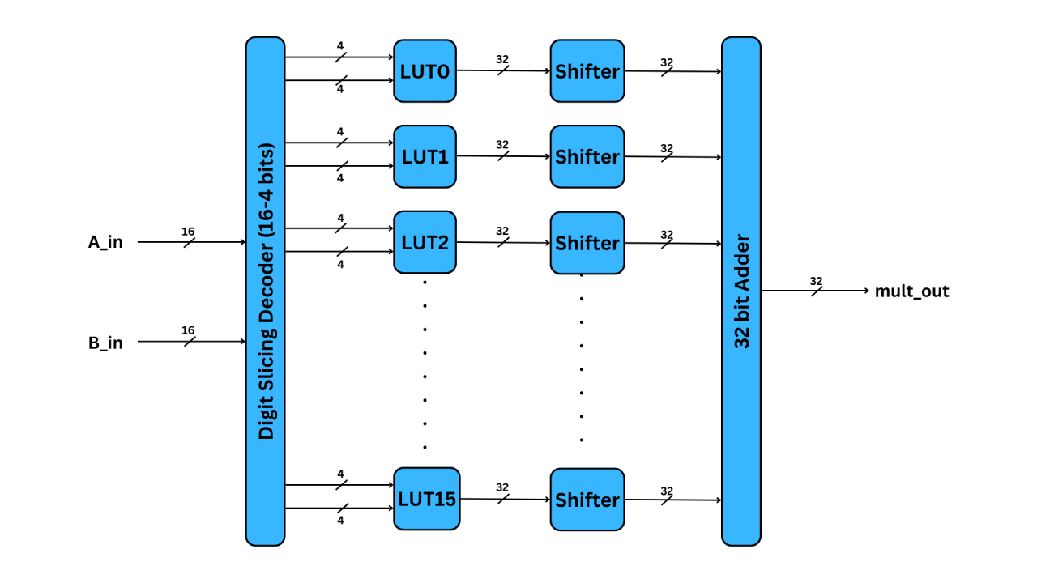}
    \caption{Bit Slicing Multiplier(BSM)}
\end{figure}

The bit slicing multiplier operates by dividing a fixed-point number into \(t\) smaller blocks, each containing \(p\) bits. These blocks independently perform \(p\)-bit multiplications using the Look-up Table (LUT) approach. Afterwards, the partial products from each block are summed to yield the final result.  The value of the input data \(x\) with length of \(B\) bits (\(x^1 ,x^2 ,x^3 ,….,x^B \)) has been represented as: 

\[ x = \sum_{j=1}^{B} {2^j x^j} \]
The sliced data \(X_k\) is represented as:
\[ X_k = \sum_{j=1}^{p} {2^j x^j} \]
Where,   
\newline \(k = 1, 2, ..., t\)\hfill \break 
\\
 After \(p\) bit multiplications, the final result  is the addition of all partial products which is represented as:
\[ F = \sum_{j=1}^{s} {X_k^j  Y_k^j} \]
Where,  
\newline \(s\) =    Number of LUTs
 \newline \(X_k\)  =     Multiplicand     
\newline \(Y_k\)    =     Multiplier\hfill \break 
\\
In our study, we've executed 16-bit multiplication utilizing 4-bit slices. We opted for 16 Look-up Tables (LUTs) to minimize latency and enhance the parallelism of the multiplication process. 

\section{General Element-wise Matrix Multiplication (GEMM) Linear Convolution:}
\begin{figure}[h]
    \centering
    \includegraphics[width=1\linewidth]{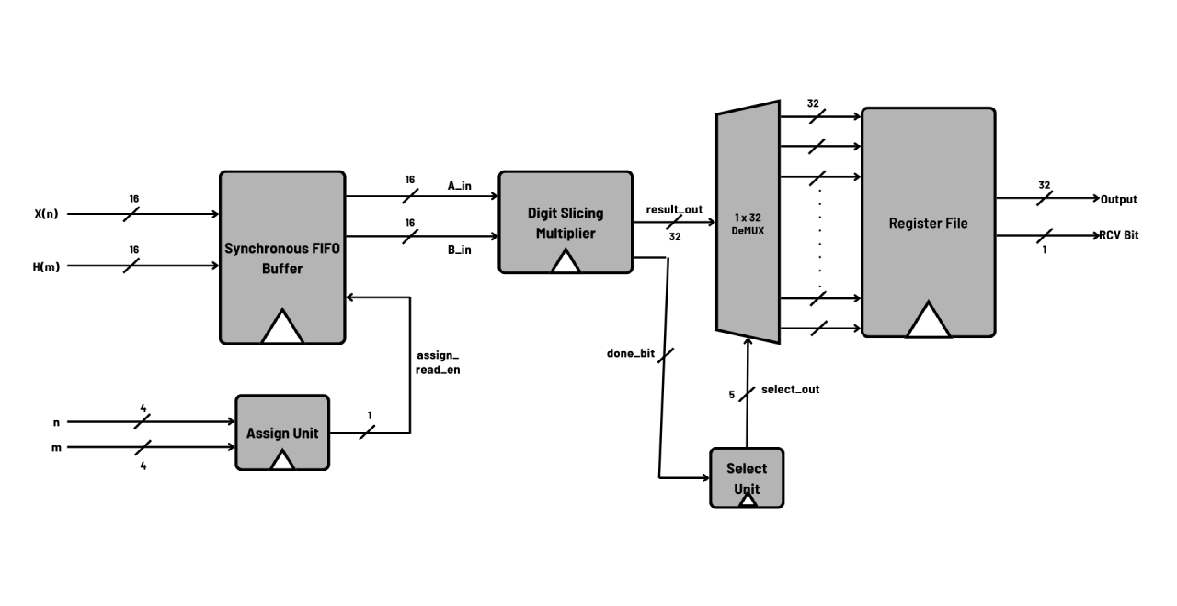}
    \caption{Convolution}
\end{figure}

The GEMM convolution process follows a systematic sequence:\hfill \break 
\\
Initially, the input array \(X(n)\) with a length of \(n\) and the kernel array \(H(m)\) with a length of \(m\) are sequentially fed into the synchronous FIFO buffer.\hfill \break 
\\
Subsequently, the kernel array \(H(m)\) is systematically positioned over the input array \(X(n)\), facilitating element-wise multiplication between the covered elements of the input array and their corresponding elements from the kernel array.   This multiplication process is facilitated by a bit slicing multiplier (BSM).\hfill \break 
\\
Following this, the 1x32 DeMUX and select unit organizes the positional arrangement of each multiplication output, which is then stored into the register file.\hfill \break 
\\
The Register file comprises 32 arrays, each with a data space of 32 bits, and is equipped with a 32-bit adder on each array. These adders compute the partial results.\hfill \break 
\\
Considering the limitations imposed by the assign unit, the maximum allowable size for the input arrays is determined to be:
\[ n,m <= {15}\]
Consequently, the maximum achievable number of final convolution outputs \(y(t)\) is:
\[ t <= {29}\]

The final convolution output \(y(t)\) can be mathematically represented as:
\[ y(t) = \sum_{m=1}^{M} {H(m) . X(t-m)} \]
The convolution output transmission is signaled by the active high state of the \(rcv\) bit.

\section{Radix-2 DIF FFT}

The N- Point DFT of an input sequence x[n] is,

\[X[k] = \sum_{n=0}^{N-1} x[n] W_n^{nk} , k=0,1,2..N-1\]
\[W_n^{nk} = e^{-j({2\pi nk}/N)}\]

Where X[k] is the output sequence and \(W_n^{nk}\) is the twiddle factor \hfill \break 

The Fast Fourier Transform (FFT) is a computational technique utilized for swiftly calculating the Discrete Fourier Transform (DFT) of a sequence or array of data points.The classic DFT algorithm operates with a computational complexity of 
\(O(N^2)\), with N representing the size of the input sequence in terms of samples.Thus it becomes computationally expensive for large values of N. The FFT algorithm was developed to compute the DFT more efficiently, reducing the time complexity to \(O(N log_2{N})\). 

\begin{figure}[h]
    \centering
    \includegraphics[width=1\linewidth]{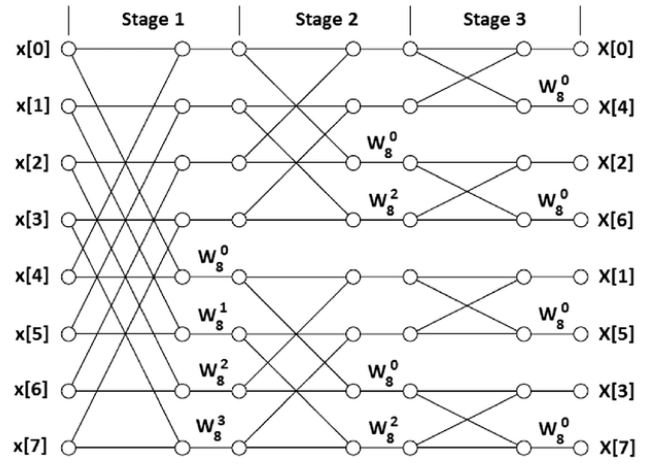}
    \caption{8-Point Radix-2 DIF FFT}
\end{figure}

The Radix-2 FFT algorithm is a specific class of FFT algorithms that efficiently computes the DFT of sequences with lengths that are powers of 2.It decomposes the DFT computation into smaller DFTs, recursively combines them using twiddle factors and butterfly operations, and produces the final frequency domain representation of the input sequence.\hfill \break 

In the R-2 DIF FFT system, the sequence x[k] is divided into 2 sub sequences :  the 1st half and the 2nd half of the sequence. Unlike the Radix-2 Decimation-in-Time (DIT) FFT algorithm, there is no requirement to reorder (shuffle) the original sequence.\hfill \break 

Splitting X(k) to odd and even terms gives ,

\[X[2k] = \sum_{n=0}^{N/2-1} [{x(n) + x(n+N/2)}]W_{N/2}^{nk}\]

\[X[2k+1] = \sum_{n=0}^{N/2-1} [ [{x(n) - x(n+N/2)}]W_{N}^{n} ]W_{N/2}^{nk}\]

Similarly, odd and even terms of each of the sequences are further reduced to implement a divide and conquer approach to implement DIF-FFT. This converts the N-point sequence into two N/2-point FFT. Each N/2-point Fast Fourier Transform is further divided into two N/4-point Fast Fourier Transform. This process continues recursively till 2-point sequences are obtained.\hfill \break 

\section{Radix-2 Single Path Delay Feedback DIF FFT}

SDF refers to the data flow structure of the FFT algorithm. In SDF FFT, there is a single path through the computation, and feedback loops are used to reuse intermediate results efficiently. It is also implemented using a pipeline architecture, where each stage of the FFT computation operates concurrently.

\begin{figure}[h]
    \centering
    \includegraphics[width=1\linewidth]{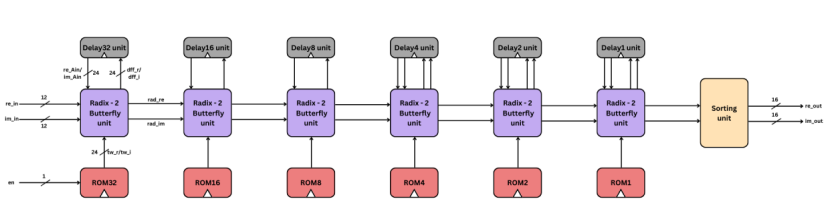}
    \caption{64-Point Radix 2 SDF DIF FFT Architecture}
\end{figure}

\textbf{Delay Unit:}
The delay unit incorporates a FIFO shift register responsible for managing and storing waiting elements. As the butterfly unit awaits the appropriate pair of elements for the computation of subsequent stage elements, the shift register dynamically adjusts its contents to ensure the correct element is presented. Each stage possesses its own delay unit tailored to its respective size. For instance, the initial stage FIFO register accommodates up to 32 elements, facilitating the waiting period until the 32nd element to initiate the next stage computation.\hfill \break 

\textbf{Butterfly Unit:}
Comprising three distinct stages, the butterfly unit executes the critical butterfly multiplication process. During the waiting stage, the unit awaits the arrival of the necessary elements, such as x[0] requiring x[32] for processing the butterfly multiplication. Upon reception, the second stage computes the summations (e.g., x[0] + x[32] and x[0] - x[32]), forwarding the results to the delay unit. Subsequently, in the third stage, multiplication between the delayed signal and the twiddle factor transpires, completing the butterfly operation.\hfill \break 

\textbf{ROM Unit:}
The ROM unit serves as the repository for the precomputed twiddle factors (Python Code calculcates the twidddle factors)  essential for multiplication operations. Upon receiving a valid signal from the preceding stage, the ROM unit initializes a counter. Based on this counter, the unit generates a state control output signal directed to the butterfly unit. During the second half of the cycle when multiplication occurs, the ROM unit furnishes the requisite twiddle factor.\hfill \break 

\textbf{Sort Unit:}
Given that the output of the DIF FFT is typically disordered, a sorting module is employed to rearrange the results systematically. This module directs the output signal to an array via a multiplexer, facilitating the sorting process. It's noteworthy that sorting an N-point FFT requires N cycles to complete.\hfill \break

The input element bits is of 12-bits whereas the twiddle factors stored are of 24 bits. So input element's size is extended to 24-bits to match the twiddle factor and again input is truncated back to 12-bits which cause inaccuracy in the output. The inaccuracy is measured in terms of SNR

\section{Results:}

In the results section of our study, we provide a comprehensive examination of power consumption and device utilization, as outlined in Tables I. These findings stem from synthesizing and simulating our developed algorithms on the FPGA Spartan-7 SP701 Evaluation Platform. Additionally, we include behavioral simulations for further insight. We have calculated the mean Signal-to-Noise Ratio (SNR) for the 64-point R2SDF FFT, taking into account design inaccuracies caused by truncation. 

\begin{figure}[h]
    \centering
    \includegraphics[width=1\linewidth]{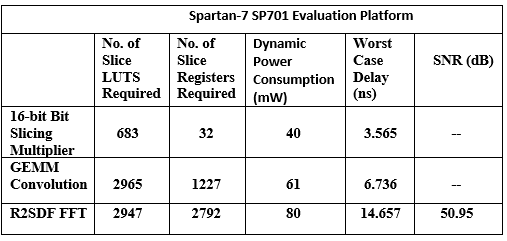}
    
\end{figure}                        
                                            \textbf{                 Table 1. }  \ Results obtained by using Bit-Slicing Multiplier
\begin{figure}[h]
    \centering
    \includegraphics[width=1\linewidth]{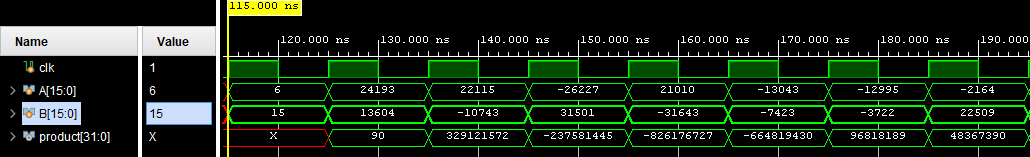}

        \caption{Behavioral Simulation Results of 16-bit Bit Slicing Multiplier}
\end{figure}\begin{figure}[h]
    \centering
    \includegraphics[width=1\linewidth]{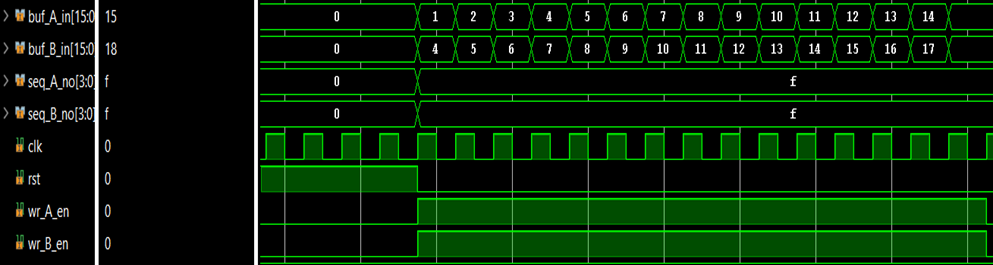}
    \includegraphics[width=1\linewidth]{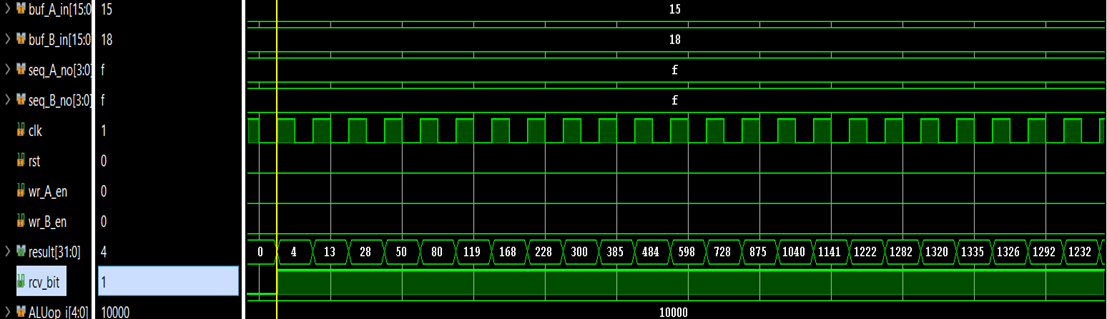}
    \caption{Behavioral Simulation Results of GEMM Convolution}
\end{figure}
\begin{figure}[h]
    \centering
    \includegraphics[width=1\linewidth]{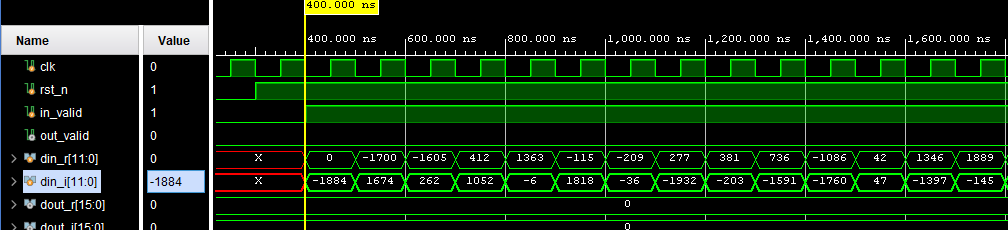}
    \includegraphics[width=1\linewidth]{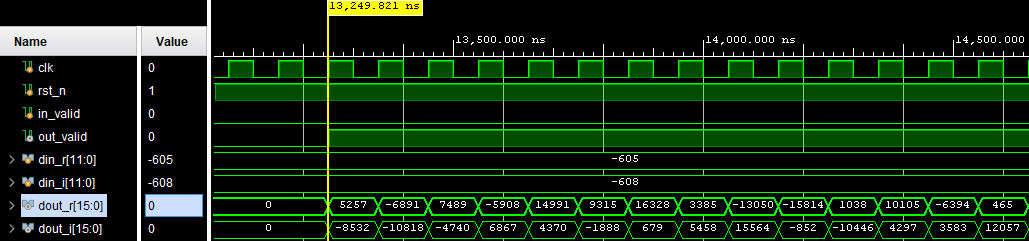}
    \caption{Behavioral Simulation Results of R2SDF DIF FFT}
\end{figure}
\vspace{\baselineskip}   

\section{Conclusion:}

In conclusion, this paper offers a thorough examination of Fast Fourier Transform (FFT) and linear convolution implementations, exploring both conventional and innovative methods, notably utilizing the Bit Slicing Multiplier (BSM) technique. Through extensive investigation and validation experiments, we have highlighted the benefits of integrating BSM technique into FFT and convolution algorithms, particularly in terms of efficiency in computation and resource utilization.

Our research emphasizes the significance of efficient signal processing methodologies, especially in scenarios where computational resources are constrained, such as real-time applications. By introducing novel concepts for FFT and convolution algorithms and demonstrating their superior performance over conventional techniques, we contribute to advancing the discussion on optimizing signal processing operations in FPGA - based systems.

\section{References:}

[1]
S. Narendran, J. Selvakumar, K. Vijayan,
Reconfigurable Superconducting FFT Processor Using Bit-Slice Block Share Processing Unit,
Microprocessors and Microsystems,
Volume 79,
2020,
103297,
ISSN 0141-9331,
https://doi.org/10.1016/j.micpro.2020.103297.
\hfill \break

[2]
E. Kalali and R. van Leuken, "Near-Precise Parameter Approximation for Multiple Multiplications on a Single DSP Block," in IEEE Transactions on Computers, vol. 71, no. 9, pp. 2036-2047, 1 Sept. 2022, doi: 10.1109/TC.2021.3119187.

\hfill \break

[3]
X. Zhou, X. Chen, Y. He and X. Mou, "A Flexible-Channel MDF Architecture for Pipelined Radix-2 FFT," in IEEE Access, vol. 11, pp. 38023-38033, 2023, doi: 10.1109/ACCESS.2023.3263880.

\hfill \break

[4]
S. Ranganathan, R. Krishnan and H. S. Sriharsha, "Efficient hardware implementation of scalable FFT using configurable Radix-4/2," 2014 2nd International Conference on Devices, Circuits and Systems (ICDCS), Coimbatore, India, 2014, pp. 1-5, doi: 10.1109/ICDCSyst.2014.6926131. 
\hfill \break

[5]
 Ö. Özdil, M. İspir, E. Onat and A. Yıldırım, "Implementation of FPGA-based FFT convolution," IET International Conference on Radar Systems (Radar 2012), Glasgow, UK, 2012, pp. 1-4, doi: 10.1049/cp.2012.1728. keywords: {Overlap-Add;FFT Convolution;Signal Processing}
 \hfill \break

[6]
S. Jiang, Y. Zou, H. Wang and W. Li, "An FFT Accelerator Using Deeply-coupled RISC-V Instruction Set Extension for Arbitrary Number of Points," in 2023 IEEE 34th International Conference on Application-specific Systems, Architectures and Processors (ASAP), Porto, Portugal, 2023 pp. 165-171.
doi: 10.1109/ASAP57973.2023.00036
\hfill \break

[7]
 M. Wang, X. Cheng, D. Zou and Z. Wang, "FACCU: Enable Fast Accumulation for High-Speed DSP Systems," in IEEE Transactions on Circuits and Systems II: Express Briefs, vol. 69, no. 12, pp. 4634-4638, Dec. 2022, doi: 10.1109/TCSII.2022.3196398.

\hfill \break

[8]
]V. M. B, M. R. A, C. S. V and A. S. K, "Fused Fixed-Point Arithmetic Unit for Radix-4 DIT FFT Implementation," 2022 IEEE 2nd Mysore Sub Section International Conference (MysuruCon), Mysuru, India, 2022, pp. 1-6, doi: 10.1109/MysuruCon55714.2022.9972587.
\hfill \break

[9]
 Elango, Sangeetha et al. “INVESTIGATION AND VLSI IMPLEMENTATION OF LINEAR CONVOLUTION ARCHITECTURE FOR FPGA BASED SIGNAL PROCESSING APPLICATIONS.” (2018)
\hfill \break

[10]
 W. Li, A. Hu, G. Wang, N. Xu and G. He, "Low-Complexity Precision-Scalable Multiply-Accumulate Unit Architectures for Deep Neural Network Accelerators," in IEEE Transactions on Circuits and Systems II: Express Briefs, vol. 70, no. 4, pp. 1610-1614, April 2023, doi: 10.1109/TCSII.2022.3231418.

\end{document}